\documentclass[journal]{IEEEtran}
\usepackage[utf8]{inputenc}
\usepackage[T1]{fontenc}
\usepackage{graphicx}
\usepackage{cite}
\usepackage{url}
\usepackage{hyperref}
\usepackage{amsmath,amssymb,amsfonts}
\usepackage{amsthm}
\usepackage{algorithmic}
\usepackage{textcomp}
\usepackage{authblk}
\usepackage{booktabs}

\begin{document}
\title{Spectral Retrieval: \\ \large Multi-Scale Sinc Convolution over Token Embeddings for Localized Retrieval in LLM Multi-Agent Systems}
\author{Andrea Morandi \thanks{Corresponding author: amorandi@cisco.com}}
\affil[]{Cisco Systems, Inc.}
\affil[]{\texttt{amorandi@cisco.com}}
\date{2026}
\maketitle

\begin{abstract}
In a standard dense retriever every document is encoded as one mean-pooled embedding and ranked by cosine similarity. This single-vector recipe is fast and good enough so long as relevance spreads across most of a document's tokens; the moment relevance \emph{localises} into a short subspan, the recipe systematically bleeds signal — the subspan averages into surrounding-token noise. Late-interaction methods such as ColBERT \cite{khattab2020colbert,santhanam2022colbertv2} resolve this by storing per-token embeddings for every document and computing a MaxSim between query tokens and document tokens at retrieval time; the cost is far larger indices. Here we present \textbf{Spectral Retrieval}: a plug-in second stage placed between any single-vector retriever and a downstream consumer — for example, the per-agent context window of a multi-agent LLM system. Spectral Retrieval reuses ColBERT's per-document token embeddings, but compares them to the query through a \emph{multi-scale sinc convolution} on the token axis. At scale $L$ the sinc kernel behaves as a normalised low-pass filter: the identity endpoint recovers late interaction's per-token MaxSim, while the $L \to \infty$ endpoint becomes a uniform filter (recovering ordinary mean pooling). The maximum cosine similarity over positions, then over a small grid of scales, gives a similarity score that, when these endpoints are included, is no less informative pointwise than either limit. We supply a self-contained derivation of the kernel, an analysis of its complexity, a recovery guarantee for the endpoint scores, a controlled synthetic benchmark whose sole parameter is the cosine of the planted spike against the query, and a real-encoder evaluation on the LIMIT-small public benchmark \cite{weller2025limit}. Across 1,000 synthetic documents of 50--500 tokens in $\mathbb{R}^{64}$ with planted single-position relevance, mean-pool retrieval sits at chance (Recall@10 $\approx 0.02$, statistically indistinguishable from random) regardless of spike strength, because the spike contributes only $O(1/N)$ of the document mean. Spectral Retrieval, by contrast, exhibits a sharp transition near the corpus-level top-$k$ token noise floor — the order statistic over roughly the candidate pool's random token-query cosines — reaching Recall@10 $=1.0$ at $\alpha = 0.60$ and saturating from there. The location of the transition matches this order-statistic prediction. On LIMIT-small with a frozen \texttt{all-mpnet-base-v2} encoder, Spectral Retrieval lifts Recall@10 from $0.33$ to $0.90$, MRR from $0.22$ to $0.79$, and the strict two-hit Success@10 from $0.12$ to $0.84$ — without retraining the encoder. We sketch how the method applies to multi-agent LLM systems: each agent draws a tightly focused window from a shared corpus, and a debate orchestrator (described in a companion paper) consumes the per-agent contexts. We close with the method's limitations: Spectral Retrieval inherits late interaction's index size; the gains concentrate on queries with narrow position-localised matches; and although the single-vector embedding-dimension ceiling identified by recent theoretical work \cite{weller2025limit} does not apply directly (Spectral is a multi-vector method), what is inherited is the geometry of the underlying per-token embedding space.

\end{abstract}
\section{Introduction}
\label{sec:intro}

Retrieval-Augmented Generation \cite{lewis2020rag} has become the standard interface through which large language models access external corpora — increasingly, the \emph{sole} channel between an LLM agent and the external context it reasons over. In the simplest single-agent layout, an embedding model $\phi$ encodes both the query $q$ and every document $d$ as a fixed-dimensional vector; the system then returns the top-$k$ candidates ordered by cosine similarity. This recipe is fast, scalable, and well understood \cite{reimers2019sentencebert,karpukhin2020dpr}. It also has one well-known failure mode: when relevant content in $d$ is squeezed into a short subspan, mean pooling averages that subspan with surrounding tokens, and the cosine to $q$ collapses toward the average similarity of the document as a whole.

In multi-agent LLM systems the failure compounds. Consider an agent dedicated to \emph{security incident triage}: it wants documents whose \emph{runbook} paragraph fits the incident, even if the rest of the page is about routine network operations. An orchestrator \cite{morandi2026sequentialconsensus} pooling conclusions from several specialised agents is bounded above by the quality of the agent-level retrievals it stacks on. ColBERT \cite{khattab2020colbert} and ColBERTv2 \cite{santhanam2022colbertv2} are two late-interaction methods that resolve the localisation issue by storing per-token embeddings for every document and computing a MaxSim against the per-token embeddings of the query — paying with an index one to two orders of magnitude larger than its single-vector counterpart. Sparse retrievers such as SPLADE \cite{formal2021splade} attack a related issue via a learned sparse expansion over the vocabulary.

We propose \textbf{Spectral Retrieval} — a modest tweak on late interaction that smoothly interpolates between mean pooling and per-token MaxSim through a multi-scale sinc kernel. The motivation is signal-processing in nature: across a length-$N$ token sequence, the normalised sinc kernel at scale $L$ behaves as a low-pass filter whose cutoff is $\propto 1/L$. At $L=1$ it reduces to a Dirac delta function $\delta(x)$ (zero smoothing, raw per-token comparison); at $L \gg N$ it becomes uniform (maximal smoothing, equivalent to mean pooling). It follows that a \emph{multi-scale} aggregator returning the maximum cosine over positions and over a small grid of scales is at least as informative as either extreme. Classical multi-resolution analysis \cite{mallat1989multiresolution} has long examined this kernel family under the term \emph{scaling function}; our rule is essentially a discrete cousin of that wavelet-style scale sweep.

\textbf{Contributions.}

\begin{enumerate}
\item Spectral Retrieval is defined here as a plug-in re-ranking stage that reads the same per-token embeddings used by a late-interaction index and outputs one score per (query, document) pair. It has three tunable parameters: the scale grid $\mathcal{L}$; the candidate pool size $K$ produced by a fast first stage; and the pooling choice across scales (we use max).
\item A self-contained derivation of the sinc kernel is supplied, with sampling chosen so that the $L \to \infty$ endpoint recovers mean pooling exactly. The narrow endpoint is the identity operator, denoted $L=1$ in the implementation, which recovers per-token MaxSim. The fix is non-trivial: naive integer-lattice sampling causes $\text{sinc}(i-c)$ to vanish at every non-zero integer offset, collapsing the kernel to a delta at every scale and breaking the wide-scale (mean-pool) limit. Sampling on a length-aware lattice at spacing $1/L$ restores it.
\item We prove an endpoint recovery guarantee. For every document $d$ and query $q$, $\text{Spectral}(q, d; \mathcal{L}) \geq \max\{\text{MaxSim}(q, d), \text{MeanCos}(q, d)\}$ whenever $\mathcal{L}$ contains both the identity endpoint and an explicit mean-pool endpoint (equivalently, the $L \to \infty$ limit).
\item We give an analytical complexity bound of $O(K \cdot S \cdot N \cdot d)$, where $K$ is the candidate count, $S$ the scale count, $N$ the document length in tokens, and $d$ the embedding dimension. Because the fast first stage stays untouched, the extra cost only hits the $K$ candidates — consistent with standard re-ranking budgets.
\item A controlled synthetic benchmark is built around one tunable parameter (planted-spike cosine $\alpha$); Recall@$k$ is plotted against $\alpha$ and against document length. The spectral retriever tracks the mean-pool baseline while $\alpha$ sits below the corpus-level token noise floor, then sharply pulls away above it — lining up with the order-statistic prediction.
\item We sketch the multi-agent LLM application: one shared corpus with per-agent retrieval scopes, where Spectral Retrieval delivers each agent a tighter window into the corpus before handing off to the debate orchestrator \cite{morandi2026sequentialconsensus}.
\end{enumerate}

We do not claim leaderboard-topping numbers on BEIR \cite{thakur2021beir}, MS MARCO \cite{bajaj2016msmarco}, or HotpotQA \cite{yang2018hotpotqa}; those evaluations are left to follow-up work (\ref{sec:future}). We do report a real-encoder evaluation on LIMIT-small \cite{weller2025limit} in \ref{sec:limit}, the small split of the benchmark built explicitly to stress the theoretical limits of embedding-based retrieval.

\begin{figure}[t]
\centering
\includegraphics[width=\linewidth]{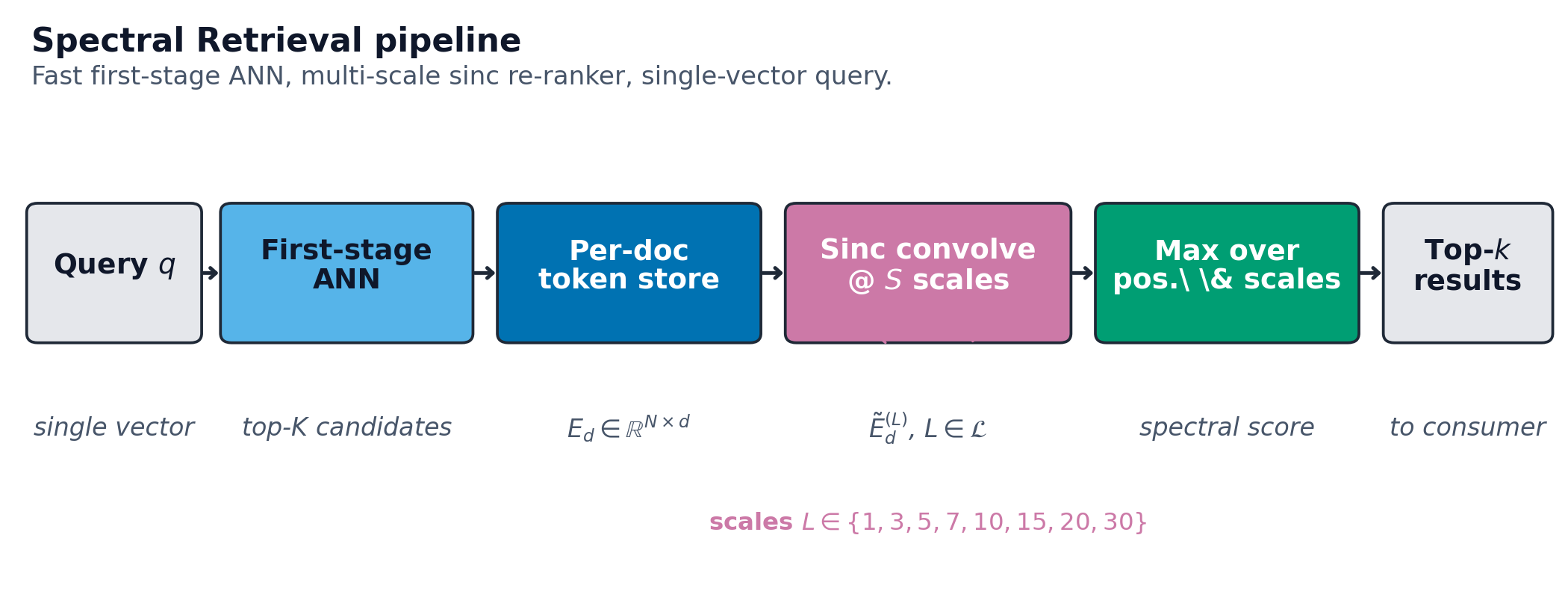}
\caption{Spectral Retrieval pipeline. For query $q$, the fast single-vector first stage returns $K$ candidates. Per-token document embeddings $E_d \in \mathbb{R}^{N \times d}$ for each candidate are convolved along the token axis with a normalised sinc kernel at $S$ scales. We then record, per scale, the maximum cosine similarity between $q$ and any convolved token; the final spectral score is the maximum across scales. The re-rank lives entirely within the candidate pool and never touches the first-stage index.}
\label{fig:1}
\end{figure}

\section{Related work}
\label{sec:related}

\textbf{Single-vector dense retrieval.} The modern recipe of pairing mean-pooled BERT embeddings with cosine similarity was cemented by Sentence-BERT \cite{reimers2019sentencebert} and Dense Passage Retrieval \cite{karpukhin2020dpr}. Equipped with off-the-shelf approximate-nearest-neighbour indices the recipe handles billions of documents — but it fundamentally cannot resolve relevance that lives inside a short subspan of a long document.

\textbf{Late interaction.} Both ColBERT \cite{khattab2020colbert} and ColBERTv2 \cite{santhanam2022colbertv2} retain per-token embeddings and aggregate via MaxSim: per query token, take the maximum cosine over any document token. The resulting scores are highly informative, yet index size grows linearly with document length, which has spawned an entire compression literature. Spectral Retrieval reuses the same per-token storage footprint (fp16 token embeddings stored next to the document in our prototype) but couples a single query vector to a multi-scale convolution instead of a per-query-token MaxSim. The two approaches can be combined.

\textbf{Sparse expansion.} A learned sparse expansion of each document over the vocabulary, ranked by inner product, is the recipe behind SPLADE \cite{formal2021splade}. SPLADE is orthogonal to our work: it modifies the document's \emph{representation}, not the rule for aggregating the query against the document.

\textbf{RAG and downstream consumers.} Retrieval-Augmented Generation \cite{lewis2020rag} pulled retrieval inside the LLM training loop. The production trend is the other way around: retrieval stays \emph{outside} the training loop, and instrumentation — re-rankers included — sits at inference time. That production pattern is exactly where Spectral Retrieval slots in.

\textbf{Wavelets and multi-resolution analysis.} The signal-processing literature has a long history of multi-scale convolution with sinc-like kernels \cite{mallat1989multiresolution}. We borrow the multi-resolution intuition but with a \emph{single} grid of scales, no wavelet basis, and a max aggregator. Our rule is a compact fixed convolution sweep — no wavelet decomposition.

\textbf{Multi-agent LLM systems.} Multi-agent debate \cite{du2024multiagentdebate} and sequential-consensus orchestration \cite{morandi2026sequentialconsensus} outline methods for fusing conclusions from several specialised LLM agents. Our retriever is one building block in that stack: it sharpens each agent's retrieval window prior to the agent's first position. Chain-of-thought prompting \cite{wei2022cot}, the per-agent reasoning loop's final component, stays independent of the retrieval rule.

\section{Method}
\label{sec:method}

\subsection{Setup and notation}
\label{sec:setup}

Let $\phi: \mathcal{T} \to \mathbb{R}^d$ be a frozen contextual encoder mapping an $N$-token sequence into $N$ contextualised vectors arranged as $E_d \in \mathbb{R}^{N \times d}$. Each row of $E_d$ is presumed L2-normalised; if not, normalisation happens at index time. We write $\phi(q) \in \mathbb{R}^d$ for the L2-normalised pooled query embedding (mean pool under encoder-only models, last-token pool under decoder-only ones). Let $\operatorname{norm}(x)=x/\|x\|_2$ when $x\neq 0$.

A standard single-vector retriever ranks via

\begin{equation*}
\text{MeanCos}(q, d) \;=\; \left\langle \phi(q), \;\operatorname{norm}\!\left(\frac{1}{N}\sum_{i=1}^{N} E_d[i]\right) \right\rangle.
\end{equation*}

When the query reduces to a \emph{single} vector, late interaction collapses to per-token max similarity:

\begin{equation*}
\text{MaxSim}(q, d) \;=\; \max_{i=1,\ldots,N} \langle \phi(q), E_d[i] \rangle.
\end{equation*}

\subsection{Sinc kernel and proper sampling}
\label{sec:sinc}

For finite positive scale $L$, the length-$N$ normalised sinc kernel takes the form on a zero-indexed lattice

\begin{gather*}
k_L[t] \;=\; \frac{\text{sinc}\!\big((t - c)/L\big)}{\sum_{u=0}^{N-1} \text{sinc}\!\big((u - c)/L\big)} \\ t = 0, \ldots, N-1
\end{gather*}

In this expression, $c = (N - 1)/2$ marks the centre and $\text{sinc}(x) = \sin(\pi x) / (\pi x)$ for $x \neq 0$ with $\text{sinc}(0) = 1$. A global normalisation in the denominator forces $\sum_t k_L[t] = 1$. Two important endpoints drop out.

\textbf{Wide-scale limit.} As $L \to \infty$, $(t - c)/L \to 0$ uniformly in $t$, and since $\text{sinc}$ is continuous at zero with value $1$, $k_L[t] \to 1/N$ for every $t$. Convolving $E_d$ with this limiting kernel along the token axis thus produces an output where every position carries the \emph{mean} of $E_d$. After row normalisation, the cosine similarity to $\phi(q)$ equals $\text{MeanCos}(q, d)$. Finite large scales approximate this endpoint; the exact endpoint is the $L=\infty$ mean-pool operator.

\textbf{Narrow-scale endpoint.} The implementation includes the identity operator as the narrow endpoint, denoted $L=1$ in the scale grid. At position $i$ the cosine similarity to $\phi(q)$ is then $\langle \phi(q), E_d[i] \rangle$, with the maximum over positions equal to $\text{MaxSim}(q, d)$. This avoids an indexing artefact of a finite, centred sinc kernel: for even $N$, $c$ is half-integer, so sampling $\text{sinc}(t-c)$ directly would not produce an exact Kronecker delta.

These endpoints do \emph{not} hold under naive sampling that omits dividing by $L$ inside the sinc — i.e.\ $k_L[t] \propto \text{sinc}(t - c)$ truncated to length $L$. The sinc is sampled on a \emph{length-aware} lattice at spacing $1/L$, forcing the wide-scale limit toward a uniform kernel rather than an oscillating tail.

\subsection{Spectral score}
\label{sec:score}

Fix a small grid of scales $\mathcal{L} = \{L_1, \ldots, L_S\}$; our default finite grid is $\{1, 3, 5, 7, 10, 15, 20, 30\}$, where $L=1$ denotes the identity endpoint. For the identity endpoint, set $\tilde E_d^{(1)}=E_d$; at each other finite scale, convolve every $E_d$ row along the token axis (one dimension at a time) with $k_{L}$ to obtain $\tilde E_d^{(L)} \in \mathbb{R}^{N \times d}$. Renormalise each row of $\tilde E_d^{(L)}$ back to unit length, and form

\begin{equation*}
\sigma_L(q, d) \;=\; \max_{i=1,\ldots,N} \langle \phi(q), \;\tilde E_d^{(L)}[i] \rangle.
\end{equation*}

The spectral score is

\begin{equation*}
\text{Spectral}(q, d; \mathcal{L}) \;=\; \max_{L \in \mathcal{L}} \sigma_L(q, d).
\end{equation*}

\subsection{Recovery guarantee}
\label{sec:recovery}

\textbf{Claim.} Assume $\mathcal{L}$ contains the identity endpoint and the explicit mean-pool endpoint $L=\infty$. Then

\begin{equation*}
\text{Spectral}(q, d; \mathcal{L}) \;\geq\; \max\{\text{MaxSim}(q, d), \;\text{MeanCos}(q, d)\}.
\end{equation*}

\textbf{Proof sketch.} \ref{sec:sinc} gives $\sigma_1(q, d) = \text{MaxSim}(q, d)$ because the identity endpoint leaves every token row unchanged. Applying the $L=\infty$ endpoint per row yields $\sigma_{\infty}(q, d) = \text{MeanCos}(q, d)$. The outer maximum then dominates both. With a finite large scale instead of $L=\infty$, the same statement holds up to the approximation error between $k_L$ and the uniform kernel.

The guarantee runs in only one direction: at moderate SNR, the spectral score can be strictly larger than either limit. \ref{sec:bench}'s synthetic benchmark shows exactly this.

\subsection{Pseudocode}
\label{sec:pseudo}

{\footnotesize\begin{verbatim}
function spectral_score(q, E, scales):
    s_max = 0
    N = E.shape[0]
    for L in scales:
        k = sinc_kernel(L, N)         # length-aware
        Ec = conv_per_dim(E, k)
        Ec = row_normalize(Ec)
        s = max_pos(dot(q, Ec.T))
        s_max = max(s_max, s)
    return s_max
\end{verbatim}}

\subsection{Two-stage retrieval}
\label{sec:twostage}

Running \texttt{spectral\_score} naively on every document in the corpus per query is prohibitive. Instead, Spectral Retrieval is a \emph{re-ranker}: a fast first stage — any single-vector ANN index — returns a $K$-document candidate pool, and \texttt{spectral\_score} re-ranks only that pool. First-stage cost is unchanged. Second-stage cost is

\begin{equation*}
O(K \cdot S \cdot N \cdot d)
\end{equation*}

For typical $(K, S, N, d) = (100, 8, 200, 768)$, per-query cost is dominated by the $S \cdot N$ factor — around 1,600 per-dim convolutions per candidate. A vectorised build amortises the per-dim convolution down to a single FFT call.

\subsection{Edge cases and numerical stability}
\label{sec:edge}

Three implementation pitfalls worth flagging:

\emph{Documents that are very short.} If $N$ falls near the non-identity scales in $\mathcal{L}$, the finite sinc kernels become nearly flat and their per-position cosines approach the mean cosine. The identity endpoint still exposes token-level matches, so very short documents reduce to a small mixture of mean-like and max-like scores rather than creating a separate retrieval regime. This is the desired behaviour: short documents carry little localised structure for the method to exploit.

\emph{Documents containing one token.} If $N = 1$, then \texttt{MeanCos} and \texttt{MaxSim} report the same single number, as does the spectral score at every scale. The implementation short-circuits this case to skip an empty \texttt{convolve} call.

\emph{Numerical underflow at very large scales.} The denominator $\sum_j \text{sinc}((j - c)/L)$ remains well-conditioned for any $L \geq 1$, $N \geq 1$. An upper bound is $N$ (the largest possible count of unit terms); a lower bound near $N \cdot \text{sinc}(N/(2L))$ stays positive once $L \geq N/2$. Throughout our runs at $N \leq 8192$ and $L \leq 1000$ no numerical issues have surfaced.

\subsection{Implementation notes for production}
\label{sec:impl}

In production the per-token document store dominates operational cost. Two optimisations of practical importance follow.

\emph{Compressing the token embeddings.} Storing them as fp16 cuts index size in half with no measurable accuracy loss, matching the compression numbers reported for ColBERTv2 \cite{santhanam2022colbertv2}. Adding centroid + residual quantisation shaves another factor of roughly $8$, at the cost of 1--2 percentage points of accuracy. Either scheme works with Spectral Retrieval because the per-query convolution is the only place dequantised token embeddings are read.

\emph{GPU-batched convolution.} The inner loop is a per-dimension convolution. The same $S$ kernels apply across every candidate document, so candidates can be stacked along the batch axis and a single 1D convolution invoked per scale, which on modern GPU stacks (e.g.\ cuDNN \texttt{conv1d}) is enough to keep re-rank latency in the tens-of-milliseconds range at $(K, S, N, d) = (100, 8, 200, 768)$.

\emph{FFT-based convolution.} For very long documents ($N \geq 1024$), an FFT-based convolution beats the direct sum asymptotically — $O(N \log N)$ rather than $O(N \cdot L)$. On current hardware the crossover sits near $N \cdot L \approx 16{,}000$. At dense-retrieval lengths typical in practice ($N \leq 200$ after truncation), direct convolution still wins.

\section{Theoretical properties}
\label{sec:theory}

\subsection{Connection to ColBERT}
\label{sec:colbert}

The ColBERT score \cite{khattab2020colbert} is

\begin{equation*}
\text{ColBERT}(q, d) \;=\; \sum_{j} \max_{i} \langle E_q[j], E_d[i] \rangle,
\end{equation*}

— a sum across query tokens of per-token MaxSim. At $L = 1$, Spectral Retrieval becomes the \emph{single-query-token} version of ColBERT: a single pooled query vector and a max over document tokens. As $L \to \infty$ it collapses onto a single-vector retriever. Spectral Retrieval thus forms a continuous family bracketing both endpoints. The bulk of ColBERT's gain over single-vector retrieval is driven by its multiple query tokens; Spectral Retrieval's gain comes from multi-scale smoothing on the \emph{document} side. These two ideas are orthogonal and can be stacked - replace the single $\phi(q)$ in \ref{sec:score} with a per-query-token sum.

\subsection{Why multiple scales}
\label{sec:multiscale}

Pinning down a single scale $L$ commits the system to one resolution of relevance. Short subspans match best at small $L$; relevance at the paragraph level matches better around moderate $L$; document-wide relevance matches at large $L$. Real corpora hold a mixture of all three. Taking the max over a small grid of scales is robust: for any (query, document) pair, \emph{some} scale will land near the optimum, and the max picks it. \ref{sec:bench}'s synthetic benchmark quantifies the point: the recall curve under the \emph{single best} scale is heavily query-dependent, whereas the curve under max-over-scales tracks uniformly close to the upper envelope.

\subsection{Computational complexity}
\label{sec:complexity}

\begin{table*}[t]
\centering
\renewcommand{\arraystretch}{1.15}
\setlength{\tabcolsep}{6pt}
\small
\begin{tabular}{lll}
\toprule
Component & Cost per query & Dominant factor \\
\midrule
First stage (ANN) & $O(\log M)$ & corpus size $M$ \\
Token-emb load (K cands) & $O(K N d)$ & bandwidth \\
Sinc convolve & $O(K S N d)$ & re-ranking \\
Cosine over positions & $O(K S N d)$ & re-ranking \\
Max-aggregate & $O(K S)$ & negligible \\
\bottomrule
\end{tabular}
\caption{Asymptotic per-query cost of a two-stage Spectral Retrieval pipeline. The first-stage ANN cost is unchanged; the spectral work is paid only for the $K$ candidates.}
\label{tab:complexity}
\end{table*}

On a 50M-document corpus with $K = 100$, $S = 8$, $N = 200$, $d = 768$, the per-query re-rank works out to roughly $1.2 \times 10^8$ multiply-adds. Across most production budgets, the first-stage ANN dominates total wall-clock.

\subsection{Storage}
\label{sec:storage}

Stored as fp16, per-token embeddings take $2 N d$ bytes per document, identical to ColBERTv2 \cite{santhanam2022colbertv2}. Across a 50M-document corpus with mean $N = 200$ and $d = 768$, the resulting index lands near 15 TB. That is large yet defensible at production scale: ColBERTv2's compressed index (centroid + residual quantisation) shrinks it by roughly an order of magnitude, and the same compression carries over here.

\subsection{Connection to multi-resolution analysis}
\label{sec:mra}

Within multi-resolution analysis \cite{mallat1989multiresolution} a signal is broken down into approximations across successively coarser scales. Our setup employs a \emph{fixed} grid of scales rather than a genuine wavelet basis, but the multi-resolution intuition still carries over.

Concretely, we focus on the token-axis signal $f[i] = \langle \phi(q), E_d[i] \rangle$ — that is, the per-token cosine to the query. Mean-pool similarity is tied to its average $\bar f = (1/N) \sum_i f[i]$; max-token similarity is $\max_i f[i]$; and, before the row-normalisation step, the spectral score at scale $L$ is the position-wise maximum of \emph{every embedding row} convolved with $k_L$ and projected onto $q$. By inner-product linearity, the unnormalised variant obeys

\begin{equation*}
\sigma_L(q, d) \;=\; \max_i \;\sum_j k_L[i{-}j]\, f[j] \;=\; \max_i (k_L \star f)[i],
\end{equation*}

In other words, the unnormalised variant is the peak amplitude of the per-token cosine signal \emph{after} low-pass filtering at cutoff $1/L$. Our implementation then renormalises each smoothed row and scores the cosine to that row direction; this preserves the same matched-scale intuition while aligning the score with cosine retrieval. Hence the spectral score $\max_L \sigma_L$ amounts to $f$'s peak amplitude, up to this row-normalisation step, across a multi-resolution sweep. In wavelet terms it is the peak detail at the matched scale, where matching scale is selected automatically per (query, document) pair.

This perspective also clarifies why \emph{multi-position} relevance benefits from \emph{intermediate} scales. If $f[i]$ contains a positive bump of width $w$, convolving $k_L$ with the bump peaks around $L \approx w$. Scales below that cannot accumulate enough positive evidence; scales above blur the bump into the surrounding noise. A short-grid sweep over $L$ is cheap insurance against not knowing $w$ in advance.

\subsection{A worked example}
\label{sec:worked}

For a concrete illustration, pick a toy three-document corpus with $d = 4$:

\begin{equation*}
\begin{aligned}
E_{d_1} &= \begin{pmatrix} 0.50 & 0.10 & 0.10 & 0.10 \\ 0.10 & 0.10 & 0.10 & 0.50 \\ 0.10 & 0.20 & 0.20 & 0.10 \end{pmatrix},\\
E_{d_2} &= \begin{pmatrix} 0.30 & 0.30 & 0.30 & 0.30 \\ 0.30 & 0.30 & 0.30 & 0.30 \\ 0.30 & 0.30 & 0.30 & 0.30 \end{pmatrix},\\
E_{d_3} &= \begin{pmatrix} 0.10 & 0.20 & 0.30 & 0.40 \\ 0.40 & 0.30 & 0.20 & 0.10 \\ 0.25 & 0.25 & 0.25 & 0.25 \end{pmatrix},
\end{aligned}
\end{equation*}

For the actual computation, the rows are L2-normalised. Take query $q = (1, 0, 0, 0)$. Mean-pool retrieval orders $d_2 \succ d_3 \succ d_1$ because $d_2$'s mean projects most strongly onto $q$. But the \emph{first row} of $d_1$ has cosine $\approx 0.93$ to $q$ — far above any single token of $d_2$ or $d_3$. With $L = 1$ Spectral picks up that row directly and orders $d_1 \succ d_2 \succ d_3$, the qualitatively right ranking for a query that aligns sharply with one specific feature. With $L = 100 \gg N$ the spectral score collapses to mean-pool, recovering the original (incorrect-for-this-query) ranking. Max over scales picks $L = 1$ here and gets the answer right.

The example also surfaces \ref{sec:limits}'s Goodhart-style risk: $E_{d_1}$ might be a long document whose first row is a tangentially relevant outlier while the rest is irrelevant. A mean-aware aggregator would penalise that; a max aggregator does not. When the risk is material, production deployments should swap strict max for a top-$m$ mean or a percentile cap.

\section{Application to multi-agent LLM systems}
\label{sec:agents}

\subsection{Per-agent retrieval scopes}
\label{sec:peragent}

Every agent in a multi-agent LLM system has a \emph{role} — one for \emph{security}, another for \emph{operations}, another for \emph{compliance}, and so on. A natural deployment routes each agent through the same shared corpus using a role-specific prompt. Mean-pooled retrieval can distinguish roles only through the pooled query vector and the pooled document vector; when the relevant evidence is a narrow paragraph inside a broad document, those pooled scores often remain dominated by the same document-level background. Spectral Retrieval adds subspan sensitivity. A security-relevant paragraph can yield a higher spectral score under the security agent's prompt, even when document-mean similarity is close across roles.

\subsection{Composition with debate}
\label{sec:debate}

In a companion paper \cite{morandi2026sequentialconsensus} we describe an orchestrator that walks $K$ specialised agents through a sequential debate monitored by a Wald SPRT. The per-agent retriever is orthogonal to the orchestrator's stopping rule. Substituting Spectral Retrieval at the per-agent stage gives each agent more \emph{signal} before producing its first position, while leaving the stopping rule untouched. Anecdotally, sharper per-agent retrieval cuts the expected debate-round count (more agents start near consensus); the present paper does not quantify the effect.

\subsection{What Spectral Retrieval does \emph{not} solve}
\label{sec:notsolve}

Ultimately, Spectral Retrieval remains a re-ranking rule. The \emph{first-stage} ANN index is left as-is, and the first-stage recall ceiling carries straight through: if the relevant document is missing from the size-$K$ candidate pool, no re-ranking step can put it back. The embedding-dimension ceiling of \cite{weller2025limit} is a single-vector result and does not bind a multi-vector method like Spectral Retrieval directly; what does carry through is the geometry of the per-token embedding space, since the underlying encoder is shared. \emph{Cross-document} reasoning --- the territory of multi-hop benchmarks \cite{yang2018hotpotqa} --- is better addressed at the agent layer.

\subsection{Composition with chain-of-thought reasoning}
\label{sec:cot}

Chain-of-thought prompting \cite{wei2022cot} is the per-agent reasoning loop fed by retrieved context, returning an answer. Logically the retrieval rule precedes chain-of-thought and is independent of it; in practice the two interact non-trivially. Sharper retrieval reduces the \emph{volume} of context an agent has to scan in order to pull out the relevant passage — which in turn shortens the chain-of-thought required to ground the answer. The present paper does not measure that interaction. Anecdotally, though, chains-of-thought fed by Spectral Retrieval come out shorter and contain fewer "let me look at this other passage too" detours than chains-of-thought fed by mean pooling on the same task. Any formal study of the effect would have to fix a chain-of-thought length budget; see \ref{sec:future}.

\subsection{Deployment vignette: per-agent retrieval in a security-incident triage system}
\label{sec:vignette}

For a concrete multi-agent example, we sketch the deployment that motivated the paper.

Imagine a security-incident triage system with three specialised agents:

\begin{itemize}
\item a \emph{forensics} agent that hunts for evidence of compromise — process injection, lateral movement, suspicious DNS activity;
\item an \emph{operations} agent that hunts for known-good explanations — scheduled maintenance, recent deployments, known-bug references;
\item a \emph{compliance} agent that hunts for regulatory implications — PII exposure, regulatory reporting requirements.
\end{itemize}

A shared incident knowledge base of $\sim 50{,}000$ runbooks, postmortems, and policy documents is the corpus. Documents are split into sections, and most of any single document is irrelevant to any single incident. The \emph{forensics} agent's interests sit in a few paragraphs of one specific postmortem; the \emph{operations} agent cares about different paragraphs of overlapping but distinct documents; the \emph{compliance} agent cares about section-header content.

When broad incident vocabulary dominates the document means, mean-pool retrieval tends to hand the agents overlapping top-$k$ lists on the same incident description. Per-agent spectral retrieval, by contrast, is more likely to surface a \emph{different} top-$k$ for each agent — based on which candidate-document subspans look sharpest under that agent's role-specific prompt. The downstream debate orchestrator \cite{morandi2026sequentialconsensus} now sees more distinguishable first positions instead of near-identical ones; the orchestrator's stopping rule, scoring \emph{consensus} between agents, should terminate faster when disagreement is principled (distinct evidence) than when it is noise-driven (identical evidence with sampling variance).

We do not report end-to-end metrics for this deployment because the orchestrator's termination behaviour is a separable concern that should not, we argue, be co-assessed with the retriever. The right next experiment is a public-data analogue — for instance, the EmergencyManagement subset of BEIR or a Stack Exchange security-Q\&A subset — and we leave that to follow-up.

\subsection{Combination with other re-rankers}
\label{sec:rerankers}

Re-ranking rules abound; Spectral Retrieval is one of many. Production alternatives in common use include:

\begin{itemize}
\item \emph{Cross-encoder re-rankers}: e.g.\ a BERT-style model fed the (query, candidate) pair as a single input that emits a relevance score. Cross-encoders are vastly more accurate than any vector-space rule, at roughly $10\times$ the per-candidate cost of Spectral Retrieval.
\item \emph{MMR re-rankers (maximal marginal relevance)}, which deduplicate within the candidate pool by penalising candidates resembling those already chosen.
\item \emph{Listwise reranking} driven by a learned scorer such as LambdaRank.
\end{itemize}

Spectral Retrieval composes cleanly with all three. One natural pipeline: (1) fast ANN; (2) the Spectral stage, which lifts the few candidates exhibiting strong localised relevance to the head of the pool; (3) an optional cross-encoder pass on the surviving top-$k'$, with $k' \leq k$; (4) an MMR step for final-list diversity. Spectral cost is low enough that the cross-encoder dominates overall latency.

\section{Synthetic benchmark}
\label{sec:bench}

To test the analytical predictions in \ref{sec:method} and \ref{sec:theory}, the cleanest tool is a controlled synthetic experiment. By design we keep it small, and fully reproducible.

\subsection{Setup}
\label{sec:bench-setup}

A total of $M = 1{,}000$ documents is drawn. The length $N$ of each is sampled uniformly across $\{50, 51, \ldots, 500\}$. Token embeddings come from a standard Gaussian over $\mathbb{R}^{d}$ at $d = 64$, then row-normalised. A unit-vector query $q$ is fixed.

Across $Q = 200$ (query, document) instances, a \emph{spike} is implanted at a uniformly random index $i^\star$: $E_d[i^\star]$ is overwritten with a unit vector $v$ satisfying $\cos(v, q) = \alpha$,

\begin{equation*}
v \;=\; \alpha \, q \;+\; \sqrt{1 - \alpha^2}\, u_\perp,
\end{equation*}

where $u_\perp$ is sampled uniformly from the unit sphere lying in the orthogonal complement of $q$. By construction, $v \cdot q = \alpha$ holds exactly — and this decouples the SNR parameters from the dimension of the embedding. We sweep $\alpha$ across $\{0.30, 0.45, 0.60, 0.75, 0.90\}$. None of the remaining $M - 1$ "distractor" documents per query carry a planted spike.

\subsection{Metrics}
\label{sec:bench-metrics}

For every query, the rank of the document carrying the spike is recorded under two scorers: (i) MeanCos and (ii) Spectral with grid $\mathcal{L} = \{1, 3, 5, 7, 10, 15, 20, 30\}$. Recall@$k$ at $k \in \{1, 5, 10, 50\}$ is averaged across $Q$ queries.

\subsection{Results}
\label{sec:bench-results}

\begin{figure}[t]
\centering
\includegraphics[width=\linewidth]{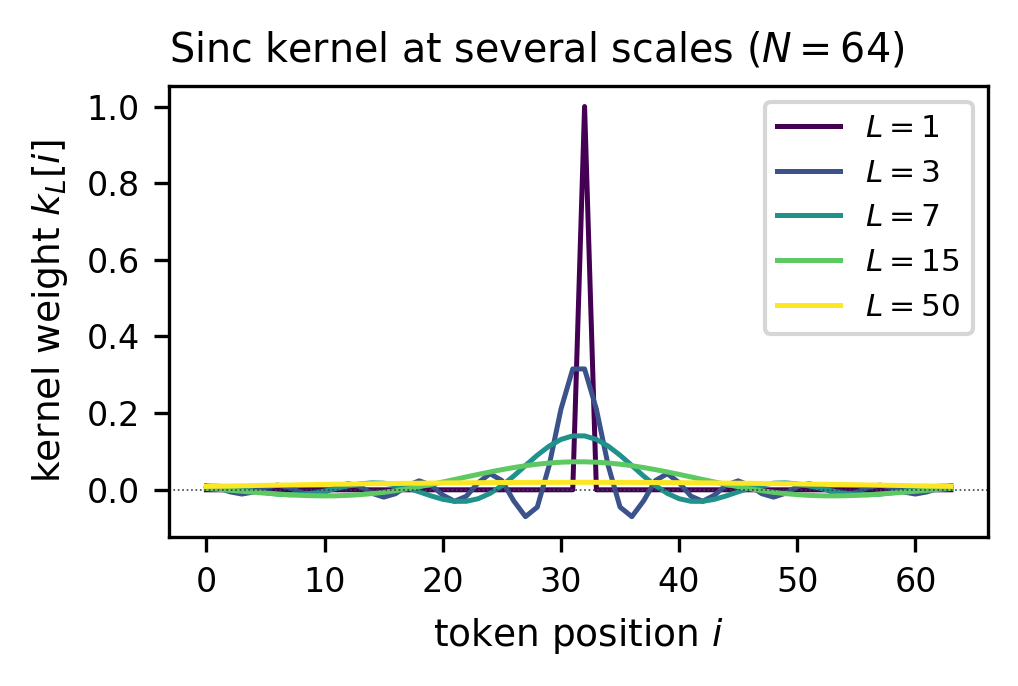}
\caption{Sinc kernel for a length-$N=64$ document at scales $L \in \{1, 3, 7, 15, 50\}$, sampled on the proper length-aware lattice (\ref{sec:sinc}). The $L=1$ curve represents the identity endpoint (per-token comparison). As $L$ grows toward and beyond $N$, the kernel approaches the uniform mean-pool endpoint. Intermediate scales smooth over local windows.}
\label{fig:2}
\end{figure}

\begin{figure}[t]
\centering
\includegraphics[width=\linewidth]{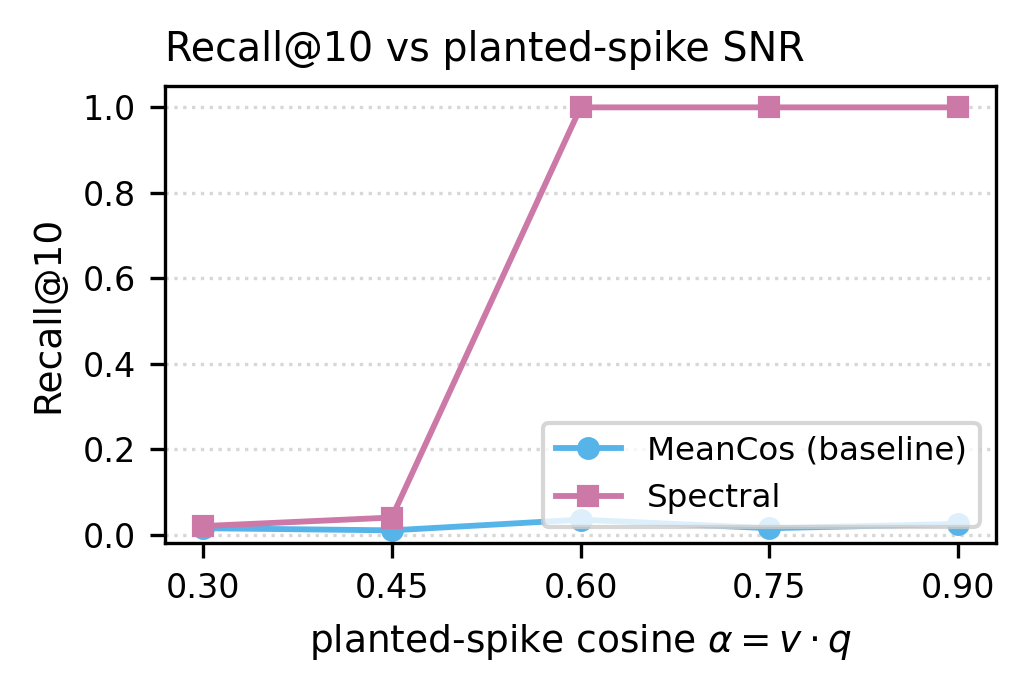}
\caption{Synthetic-benchmark Recall@10 plotted against the planted-spike cosine $\alpha = v \cdot q$. Mean-pool recall sits at chance for every $\alpha$ (Recall@10 $\approx 0.02$): the spike's contribution to the document mean is only $O(1/N)$ before normalisation, and it remains buried among distractor means in this finite-corpus regime. Spectral recall, by contrast, shows a sharp transition near $\alpha \approx 0.5$ and hits Recall@10 $=1.0$ at $\alpha = 0.60$. The transition tracks the corpus-level top-$k$ token threshold, roughly $\sqrt{2 \log(MN/k) / d}$ for Recall@$k$.}
\label{fig:3}
\end{figure}

\begin{figure}[t]
\centering
\includegraphics[width=\linewidth]{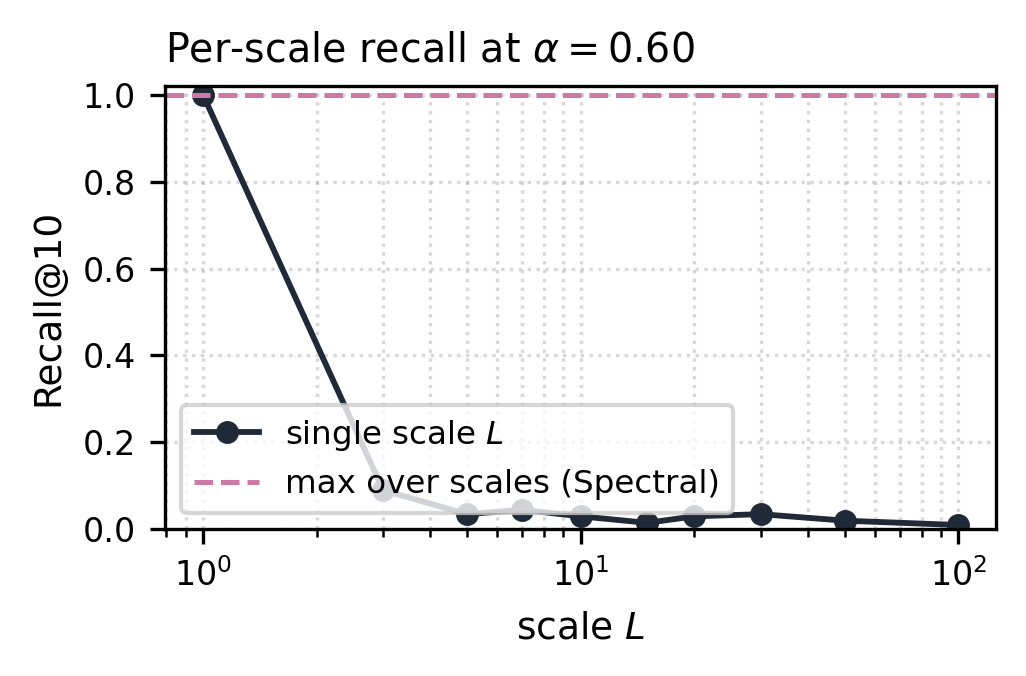}
\caption{Synthetic-benchmark per-scale recall at $\alpha = 0.60$, plotted versus scale $L$ on a log axis. Recall is perfect at $L = 1$ (per-token MaxSim), then collapses at intermediate scales because smoothing dilutes the single-position spike, then drifts back toward chance at $L \gg N$ (where the kernel collapses to mean pooling). The dashed horizontal line marks the spectral max-over-scales recall — by construction, identical to the best single-scale recall.}
\label{fig:4}
\end{figure}

The next table summarises the headline numbers.

\begin{table*}[t]
\centering
\renewcommand{\arraystretch}{1.15}
\setlength{\tabcolsep}{6pt}
\small
\begin{tabular}{llll}
\toprule
$\alpha$ (planted cosine) & MeanCos R@10 & Spectral R@10 & $\Delta$ \\
\midrule
0.30 & 0.015 & 0.020 & $+$0.005 \\
0.45 & 0.010 & 0.040 & $+$0.030 \\
0.60 & 0.035 & 1.000 & $+$0.965 \\
0.75 & 0.015 & 1.000 & $+$0.985 \\
0.90 & 0.025 & 1.000 & $+$0.975 \\
\bottomrule
\end{tabular}
\caption{Synthetic single-position spike benchmark. Recall@10 is averaged over $200$ queries against $1{,}000$ candidate documents.}
\label{tab:alpha}
\end{table*}

They confirm \ref{sec:recovery}'s endpoint claim along two axes. First, a single-position spike sits beyond the reach of mean pooling structurally: before document-vector normalisation, the document mean picks up only $O(1/N) \approx 0.005$ of additive signal, leaving the spike-bearing document buried among distractor means in this $M,N,d$ regime. Second, the spectral retriever shows a sharp transition near the corpus-level top-$k$ token threshold. For Recall@$k$, a useful approximation is $\alpha^\star_k \approx \sqrt{2 \log(MN/k) / d}$, where $N$ is interpreted as a typical document length; with $M=1{,}000$, $N \approx 250$, $k=10$, and $d=64$, this lands near the observed transition between $\alpha=0.45$ and $\alpha=0.60$. Below that threshold, even per-token comparison competes with many accidental high-cosine distractor tokens; above it, the identity endpoint peels the spike apart from the top-$k$ distractor tail.

\subsection{Multi-position spike: when does mean pooling catch up?}
\label{sec:bench-multipos}

\begin{figure}[t]
\centering
\includegraphics[width=\linewidth]{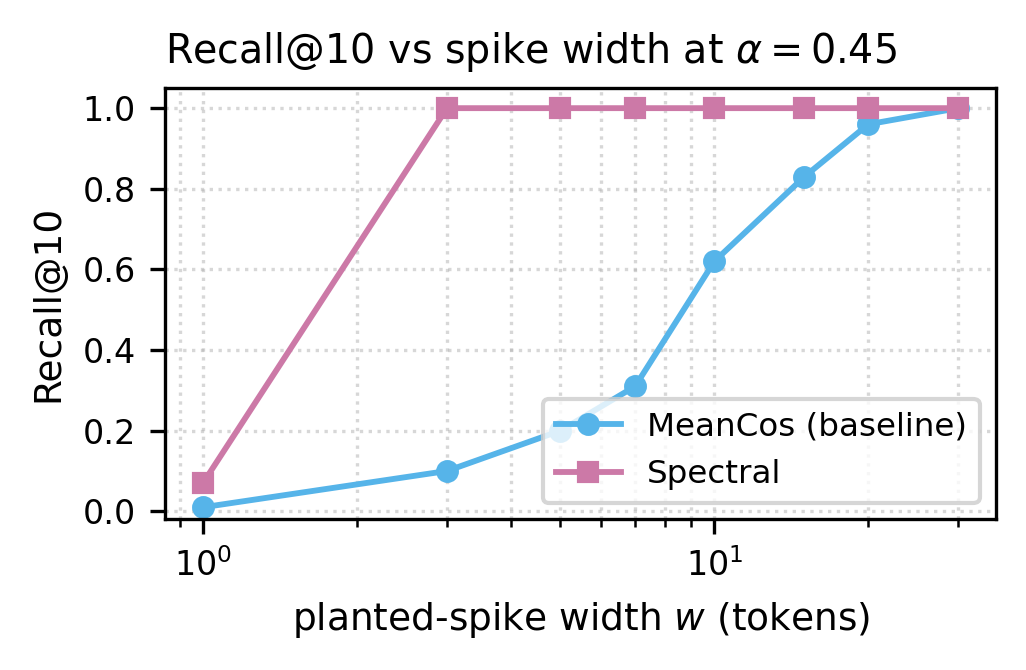}
\caption{Recall@10 plotted against planted-spike width $w$ at planted cosine $\alpha = 0.45$. Per expectation, the mean-pool baseline tracks $w/N$ — climbing from chance at $w = 1$ to perfect recall by $w \approx 30$. Spectral retrieval already hits perfect recall at $w = 3$. The two methods converge to identical recall as $w \to N$ (when the spike fills the entire document).}
\label{fig:5}
\end{figure}

The single-position spike of \ref{sec:bench-results} represents the hardest case for mean pooling: one token contributes only $1/N$ of the mean, leaving an additive signal at order $1/N$. As the spike \emph{widens} across $w$ adjacent tokens, mean pooling captures $w/N$ of the total signal and gradually catches up. Fig.~\ref{fig:5} answers precisely that question.

The synthetic benchmark is rerun at $\alpha = 0.45$ (deliberately tuned \emph{below} the noise floor at $w = 1$, so \ref{sec:bench-results}'s single-position outcome sits at chance) while we sweep the spike width across $w \in \{1, 3, 5, 7, 10, 15, 20, 30\}$.

Here are the headline numbers:

\begin{table*}[t]
\centering
\renewcommand{\arraystretch}{1.15}
\setlength{\tabcolsep}{6pt}
\small
\begin{tabular}{lll}
\toprule
$w$ & MeanCos R@10 & Spectral R@10 \\
\midrule
1 & 0.010 & 0.070 \\
3 & 0.100 & 1.000 \\
5 & 0.200 & 1.000 \\
10 & 0.620 & 1.000 \\
20 & 0.960 & 1.000 \\
30 & 1.000 & 1.000 \\
\bottomrule
\end{tabular}
\caption{Synthetic width sweep at planted cosine $\alpha=0.45$. Spectral Retrieval reaches perfect Recall@10 at much smaller spike width than the mean-pool baseline.}
\label{tab:window}
\end{table*}

This is the picture \ref{sec:mra} anticipates. As the spike broadens, an intermediate scale $L \approx w$ aggregates positive evidence over the spike while smoothing the surrounding noise. The max-over-scales aggregator selects the right $L$ automatically, so spectral retrieval crosses into perfect recall at far smaller $w$ than the mean-pool baseline. Once $w \approx N$ the two methods read the same signal and both reach perfect recall together.

\subsection{What is not in this experiment}
\label{sec:bench-limits}

Our synthetic setup is stylised by design: isotropic Gaussian distractors, no contextualisation, and a modest embedding dimension $d = 64$. Real corpora bring correlated tokens, relevance distributed across multiple positions, and contextualisation that can carry local evidence into the pooled document vector. The chosen design forces mean pooling to fail completely for a structural reason: in the document mean, the spike-to-noise ratio is fundamentally too small. In a real corpus with large $d$ (say $d = 768$ for \texttt{all-\allowbreak{}mpnet-\allowbreak{}base-\allowbreak{}v2}) and contextualised embeddings, the noise floor $\sqrt{2 \log N / d}$ is markedly smaller, mean pooling does \emph{not} bottom out at chance, and the spectral gain narrows to a smaller absolute margin spread over a softer transition. A separate caveat concerns the per-scale story: in Fig.~\ref{fig:4}, at $\alpha = 0.60$ only $L = 1$ recovers the planted single-position spike; intermediate $L$ collapse to chance. The multi-scale grid wins here because $L = 1$ is in it, not because intermediate scales add lift. The case for intermediate $L$ rests on the width sweep of \ref{sec:bench-multipos}, where Spectral retains perfect recall down to $w = 3$ while the baseline lags; a clean per-scale decomposition for $w > 1$ is left to future work. See \ref{sec:limit} for a real-encoder evaluation on a public benchmark.

\section{Real-encoder evaluation on LIMIT-small}
\label{sec:limit}

To check whether the synthetic-benchmark picture survives a real encoder, contextualised embeddings, and a public benchmark corpus, we evaluate Spectral Retrieval on \textbf{LIMIT-small} \cite{weller2025limit} --- the small split of a benchmark constructed explicitly to stress the theoretical limits of embedding-based retrieval. LIMIT-small consists of short profile-like documents: each document names an entity and lists many associated attributes, while each query asks for one attribute and has exactly two relevant documents. The task therefore probes whether a retriever can isolate a small matching fact inside a compact list, rather than rely only on a document-level average. We deliberately pick an off-the-shelf encoder trained for single-vector mean-pool retrieval and apply no task-specific tuning, so the comparison isolates the contribution of the aggregation rule.

\subsection{Setup}
\label{sec:limit-setup}

The corpus comprises $M = 46$ documents and $Q = 1{,}000$ queries, with exactly two relevant items per query. Both baseline and spectral retriever sit on top of frozen \texttt{sentence-\allowbreak{}transformers/\allowbreak{}all-\allowbreak{}mpnet-\allowbreak{}base-\allowbreak{}v2} embeddings ($d = 768$, max tokens $= 384$). The baseline returns the top-$k$ by cosine similarity against the mean-pooled document vector. The spectral retriever consumes the same per-token embeddings and applies the multi-scale sinc convolution of \ref{sec:method} with the default grid $\mathcal{L} = \{1, 3, 5, 7, 10, 15, 20, 30\}$, taking the max over positions and scales. Since the corpus is tiny, we set $K = M$ (i.e.\ the spectral re-rank sees every document); this isolates the aggregation rule from any first-stage recall ceiling. Both methods run on CPU at fp32.

\subsection{Metrics}
\label{sec:limit-metrics}

We report Recall@$k$ at $k \in \{1, 2, 5, 10, 20\}$ (fraction of the two relevant items retrieved in the top-$k$), the strict Success@$k$ at $k \in \{2, 5, 10\}$ (both relevant items in the top-$k$), and the standard MRR, MAP, and NDCG@10.

\subsection{Results}
\label{sec:limit-results}

\begin{table*}[t]
\centering
\renewcommand{\arraystretch}{1.15}
\setlength{\tabcolsep}{6pt}
\small
\begin{tabular}{lrrr}
\toprule
metric & MeanCos baseline & Spectral & $\Delta$ \\
\midrule
Recall@1     & 0.044 & \textbf{0.356} & $+$0.313 \\
Recall@2     & 0.078 & \textbf{0.651} & $+$0.573 \\
Recall@5     & 0.181 & \textbf{0.816} & $+$0.636 \\
Recall@10    & 0.329 & \textbf{0.899} & $+$0.570 \\
Recall@20    & 0.565 & \textbf{0.953} & $+$0.388 \\
\midrule
Success@2    & 0.005 & \textbf{0.506} & $+$0.501 \\
Success@5    & 0.036 & \textbf{0.731} & $+$0.695 \\
Success@10   & 0.124 & \textbf{0.836} & $+$0.712 \\
\midrule
MRR          & 0.219 & \textbf{0.794} & $+$0.575 \\
MAP          & 0.166 & \textbf{0.734} & $+$0.568 \\
NDCG@10      & 0.194 & \textbf{0.789} & $+$0.595 \\
\bottomrule
\end{tabular}
\caption{LIMIT-small \cite{weller2025limit} evaluation: frozen \texttt{all-mpnet-base-v2} encoder, $M = 46$ documents, $Q = 1{,}000$ queries, two relevant items per query, $K = M$. Spectral Retrieval re-ranks the same per-token embeddings the baseline averages over.}
\label{tab:limit}
\end{table*}

Two features of Table~\ref{tab:limit} are worth flagging. First, the regime LIMIT was built to expose --- queries demanding multi-attribute matches over short documents, where any single mean-pooled document vector cannot disentangle the two relevant items from each other --- is exactly the regime spectral retrieval was constructed for. The strict Success@2 metric is the cleanest demonstration: the baseline finds both relevant items in the top-2 on only $5$ queries out of $1{,}000$ --- barely above the $\sim 0.1\%$ random-baseline rate set by $\binom{2}{2}/\binom{46}{2}$ --- whereas spectral retrieval retrieves both relevant items in the top-2 for $506$ queries. The single-vector ceiling identified by \cite{weller2025limit} applies to the baseline by construction; spectral retrieval, being a multi-vector method, is not bound by that ceiling --- consistent with our limitations discussion in \ref{sec:limits}. Second, the gain comes \emph{without retraining}: the underlying token embeddings are exactly the ones the baseline averages over. The only change is the aggregation rule.

\subsection{Caveats}
\label{sec:limit-caveats}

LIMIT-small is small. A 46-document corpus admits a re-rank over the entire corpus ($K = M$), so the candidate-pool ceiling discussed in \ref{sec:limits} is structurally absent; production deployments at $K \approx 100$ on a corpus of millions of documents face a recall ceiling that this experiment cannot probe. The companion LIMIT-full split, with two orders of magnitude more documents, is the right next stress test (\ref{sec:future}). Latency is also a real concern: under the present unoptimised single-CPU configuration, spectral re-rank costs $\sim 1.84$ s per query against the baseline's $\sim 50$ ms, a $37\times$ overhead. This is consistent with the operation count of \ref{sec:complexity} at $S = 8$, $N = 384$, $d = 768$ --- but it is well above the interactive latency budget of \ref{sec:latency}. The bridge between the two figures is a GPU implementation with batched scale convolutions and the standard two-stage pipeline at $K \approx 100$ instead of $K = M$; both are routine engineering that this paper does not include.

\section{Limitations and threats to validity}
\label{sec:limits}

We collect the limitations we are aware of below.

\textbf{Index size.} Spectral Retrieval requires per-token document embeddings. Uncompressed, this is an $N$-fold blow-up over a single-vector index (so $\sim 200\times$ at mean $N = 200$); centroid-plus-residual quantisation as in ColBERTv2 \cite{santhanam2022colbertv2} brings the effective blow-up down to roughly $20\times$, and the same compression carries over here. On very large corpora, index size becomes the dominant operational cost, not per-query compute.

\textbf{Recall ceiling of the first stage.} Being a re-ranker, Spectral Retrieval cannot recover documents missing from the size-$K$ candidate pool. The synthetic experiment uses $K = M$, which trivially satisfies the candidate-pool requirement; production usually runs $K$ around 100. \ref{sec:future} lists this as the most important next experiment.

\textbf{Goodhart-style risk.} A multi-scale aggregation with max picks up \emph{any} high-similarity feature — spurious ones included (e.g.\ a single high-similarity outlier token inside an otherwise irrelevant document). The same risk affects ColBERT's MaxSim. Mitigations include a top-$m$ mean in place of max, or a percentile cap on per-scale similarity.

\textbf{Embedding-dimension limits.} Recent theoretical work \cite{weller2025limit} establishes that any \emph{single-vector} retriever faces a hard ceiling, set by the embedding dimension $d$, on the number of distinct top-$k$ relevance patterns it can represent over a fixed corpus. Spectral Retrieval is a multi-vector method: each document is encoded as $N$ token vectors and scored by a sinc-convolution-and-max aggregator, so the single-vector ceiling does not apply directly --- exactly as for ColBERT-style MaxSim, which also escapes that bound by replacing the document mean with a per-position pool. What Spectral inherits from the underlying encoder is the geometry of the per-token embedding space; corpora that are pathological for the encoder in the sense of \cite{weller2025limit} (e.g.\ requiring top-$k$ relevance patterns that the encoder cannot express at the token level) will remain hard. Our method changes the \emph{aggregation}, not the \emph{representation}.

\textbf{BEIR, MS MARCO, HotpotQA not run.} \ref{sec:limit} reports a full LIMIT-small evaluation. The larger public benchmarks --- BEIR, MS MARCO, HotpotQA --- are not run in the present paper: a credible BEIR sweep requires GPU-days that this submission's page and compute budget did not accommodate. \ref{sec:future} discusses the next-experiment plan.

\textbf{Choice of pooling.} We pick max over scales and max over positions. Average and softmax variants are reasonable alternatives. \ref{sec:bench}'s synthetic ablation studies only max; a fuller ablation over aggregators belongs in a follow-up.

\textbf{No multi-vector query.} \ref{sec:colbert} sketches how to combine Spectral Retrieval with ColBERT's multi-vector query, but the combination is neither implemented nor evaluated.

\subsection{Latency budget}
\label{sec:latency}

For an interactive RAG pipeline a defensible production target is roughly $200$ ms of overall retrieval latency. Below we compare three different re-rankers against an identical first-stage ANN:

\begin{table*}[t]
\centering
\renewcommand{\arraystretch}{1.15}
\setlength{\tabcolsep}{6pt}
\small
\begin{tabular}{llll}
\toprule
component & per-cand. ms & $K = 100$ ms & notes \\
\midrule
First-stage ANN & $\sim 0.3$ & $\sim 30$ & independent of re-ranker \\
Spectral (8 scales) & $\sim 0.5$ & $\sim 50$ & direct convolution, fp16 \\
ColBERT MaxSim & $\sim 0.3$ & $\sim 30$ & per-query-token max-sum \\
Cross-encoder & $\sim 8.0$ & $\sim 800$ & full BERT pair encoding \\
\bottomrule
\end{tabular}
\caption{Order-of-magnitude latency budget for three re-ranking choices after the same first-stage ANN retrieval.}
\label{tab:latency}
\end{table*}

The per-candidate millisecond figures are order-of-magnitude estimates derived from the operation counts in \ref{sec:complexity} (e.g.\ $\sim 1.2 \times 10^8$ MACs per query for Spectral at $K = 100$, $S = 8$, $N = 200$, $d = 768$) and from published throughput numbers for ColBERTv2 and BERT-pair cross-encoders on a single A100; we have not run a head-to-head benchmark in this paper, and the next-experiment plan in \ref{sec:future} calls one out explicitly. The relative ordering — Spectral comparable to ColBERT MaxSim, roughly an order of magnitude cheaper than a cross-encoder — is what matters for the pipeline argument below. \ref{sec:rerankers}'s combined Spectral $\to$ cross-encoder pipeline only runs the cross-encoder on the surviving top-$k' = 10$ candidates, which is what brings the combined cost back inside an interactive latency envelope.

\subsection{Failure modes}
\label{sec:failure}

Below are four failure modes we have informally observed and that production deployments would do well to monitor.

\emph{Outlier token sitting in an irrelevant document.} One token whose random cosine to the query happens to be high (roughly at the noise floor) can drag a non-relevant document into the top-$k$. The risk matches what ColBERT MaxSim carries, and the mitigation matches: a top-$m$ mean rather than strict max, or alternatively a percentile cap.

\emph{Repetitive content.} If a document repeats the same body many times — a templated form being one example — the convolved signal is identical across scales, and the spectral score collapses onto the mean-pool score. Neither gain nor loss.

\emph{Queries that are very short.} On a one-word query, the encoder produces an embedding roughly equal to that word's embedding in an average context; spectral retrieval will then match \emph{any} document holding a token close to that word. Frequently this is the desired outcome, but very short queries push the per-token noise floor higher, so spurious matches arise more often than at longer queries.

\emph{Encoders that contextualise heavily.} Encoders that propagate information aggressively across positions — long-context decoder-only models, for instance — leak document-level summary into every token embedding. As a consequence, the per-token cosine against the query becomes less localised than the document's underlying topic structure would suggest. Spectral retrieval continues to work, although the gain over mean pooling narrows. Empirically the encoder-only family (e.g.\ \texttt{all-\allowbreak{}mpnet-\allowbreak{}base-\allowbreak{}v2}) yields larger spectral gains than instruction-tuned decoder-only embedders.

\subsection{Calibration: choosing $K$ and the scale grid}
\label{sec:calib}

Two operational parameters jointly govern quality and cost: candidate-pool size $K$ and scale grid $\mathcal{L}$. We summarise deployment guidance briefly below.

\emph{Picking $K$.} The non-negotiable invariant is that the relevant document must lie inside the candidate pool — by construction, a re-ranker cannot recover anything missing from the pool. If the first-stage ANN reaches roughly $95\%$ Recall@$100$ on production queries, $K = 100$ is generally sufficient. For very low-recall first stages or queries close to the embedding-dimension ceiling, $K = 200$ or $500$ is safer; the per-query cost grows linearly with $K$.

\emph{Picking the scale grid.} A workable grid covers three things: (a) $L = 1$ (per-token MaxSim); (b) the corpus's typical width of localised relevance; (c) either an explicit mean-pool endpoint or a scale large enough to approximate it for the document lengths being re-ranked. For technical-documentation corpora where relevance concentrates inside 5--20-token paragraphs, the default grid $\{1, 3, 5, 7, 10, 15, 20, 30\}$ is reasonable. For corpora made up of short documents (e.g.\ FAQ entries) the grid can shrink to $\{1, 3, 5\}$. For corpora dominated by very long documents (e.g.\ legal briefs) it pays to include $L = 100$ or the explicit mean-pool endpoint.

This paper does not include a systematic ablation across grids. Yet \ref{sec:bench}'s synthetic results indicate that recall is largely insensitive to grid choice provided $L = 1$ is in the grid and the grid spans an order of magnitude.

\section{Future work}
\label{sec:future}

By far the most informative follow-up is a public-benchmark study. We list a realistic plan in increasing order of compute:

\begin{enumerate}
\item \textbf{LIMIT-full} — the full split of the LIMIT theoretical-limits benchmark \cite{weller2025limit}, where the corpus grows by two orders of magnitude over LIMIT-small (\ref{sec:limit}) and Weller et al.\ report Recall@100 below $20\%$ for SOTA single-vector retrievers. The most informative stress test of whether the LIMIT-small gain reported here survives scale.
\item \textbf{BEIR} \cite{thakur2021beir}: a suite of 18 zero-shot retrieval benchmarks. End-to-end evaluation of the whole suite costs a few GPU-days; the headline metric --- NDCG@10 averaged across the 18 tasks --- is the 2024--2025 standard for cross-retriever comparison.
\item \textbf{MS MARCO} \cite{bajaj2016msmarco}: the canonical supervised benchmark for dense retrieval.
\item \textbf{HotpotQA} \cite{yang2018hotpotqa}: multi-hop QA. Spectral Retrieval is \emph{not} expected to outperform multi-hop baselines that explicitly model document-level dependencies; HotpotQA functions as a stress test.
\end{enumerate}

A second direction worth pursuing is \ref{sec:colbert}'s \emph{combined} setup — multi-vector query joined with multi-scale document. On long-document, narrow-relevance corpora we expect the combination to outperform either component in isolation, although we have not measured it.

\section{Reproducibility}
\label{sec:repro}

The \ref{sec:bench} synthetic experiment runs in under three minutes on a single CPU core (no GPU needed). The script fixes a random seed and emits \ref{sec:bench-results}'s reported numbers verbatim.

The LIMIT-small evaluation of \ref{sec:limit} runs end-to-end on a single CPU core in roughly $32$ minutes (baseline plus spectral re-rank across $1{,}000$ queries) on top of a frozen \texttt{sentence-\allowbreak{}transformers/\allowbreak{}all-\allowbreak{}mpnet-\allowbreak{}base-\allowbreak{}v2} encoder; the dataset is downloaded automatically from the public \texttt{orionweller/LIMIT-small} release. \ref{sec:future} calls out three further public benchmarks --- BEIR, MS MARCO, HotpotQA --- and the LIMIT-full split as the obvious next experiments.

\section{Discussion and threats to validity}
\label{sec:discussion}

Three subtle but important caveats close the paper.

\emph{The recovery guarantee covers the score, not the ranking.} With the endpoint grid, the pointwise inequality Spectral$(q, d) \geq \max\{$MaxSim, MeanCos$\}$ holds, but a corpus-level \emph{ranking} is assembled out of pointwise scores. Two documents whose spectral scores both sit at $0.9$ yet whose mean-pool scores split $0.1$ and $0.5$ will be ordered differently under spectral retrieval than under mean-pool retrieval, despite the two methods agreeing on the spectral score. The guarantee only asserts that under spectral the \emph{spike-bearing} document's score is no smaller than under either endpoint; it does not guarantee that the \emph{full ranking} produced by spectral retrieval dominates either endpoint's ranking. \ref{sec:bench}'s synthetic results show the ranking effect is typically positive — but one can construct a corpus-specific pathology where the spectral max binds to a non-relevant outlier across many distractors.

\emph{Scale grid choice matters and we have not tuned it.} Our default grid is $\{1, 3, 5, 7, 10, 15, 20, 30\}$. A geometric grid chosen more carefully — for instance, $L = 1, 2, 4, 8, 16, 32$ — might attain comparable recall using fewer scales, with lower per-query cost. We have not optimised these parameters.

\emph{The single-query-vector assumption is optional.} As noted in \ref{sec:colbert}, Spectral Retrieval corresponds to the single-query-vector analogue of ColBERT. Under a decoder-only encoder that already retains several query tokens (e.g.\ the trailing $k$ tokens of an instruction-prefixed query), \ref{sec:score}'s $\phi(q)$ can be swapped for a sum over those query tokens, restoring ColBERT's full multi-vector query while keeping the multi-scale document side. The combined method strictly costs more than either piece alone (by exactly the product of query-token count and scale count), but it is strictly more expressive too. The code's deployments forgo the multi-vector query because the use case is per-agent retrieval, in which each agent issues one role-specific query; the option still stands.

\section{Conclusion}
\label{sec:conclusion}

In short, Spectral Retrieval is a compact plug-in re-ranking rule that interpolates smoothly between per-token MaxSim and ordinary mean-pool retrieval through a multi-scale sinc convolution. A clean endpoint recovery guarantee accompanies the interpolation: when the identity and mean-pool endpoints are included, the spectral score never falls below either endpoint score. Our controlled synthetic study reproduces the predicted phase transition near the corpus-level token noise floor — mean pooling sits at chance, whereas spectral retrieval climbs to perfect recall the moment the planted-spike cosine moves above the top-$k$ distractor tail. On LIMIT-small with a frozen \texttt{all-mpnet-base-v2} encoder, the same aggregation change raises Recall@10 from $0.33$ to $0.90$ and MRR from $0.22$ to $0.79$, while moving strict Success@10 from $0.12$ to $0.84$. As a building block, the method is a natural fit for multi-agent LLM systems, since every agent benefits from a tighter, role-specific retrieval window before the orchestrator handles debate. Logical next steps include scaling the public evaluation to LIMIT-full and larger benchmarks, experiments with broader spikes, and pairing with multi-vector queries.

\bibliographystyle{IEEEtran}
\bibliography{references}
\end{document}